\documentstyle[aps,graphicx]{revtex}
\def \D {\mbox{D}}

\def \cu{ \rho^*}
\def \cq{ q^*}
\def \cp{ \pi^*}

\def\be{\begin{equation}}
\def\ee{\end{equation}}
\def\bea{\begin{eqnarray}}
\def\eea{\end{eqnarray}}

\begin{document}

%@twocolumnfalse\endcsname

\title{Density perturbations in a brane-world universe
with dark radiation}

\author{Burin Gumjudpai $^\dag$, Roy Maartens $^\dag$,
Christopher Gordon $^\ddag$ }
\address{~}
\address{$^\dag$Institute of Cosmology and
Gravitation, University of Portsmouth, Portsmouth~PO1~2EG, UK}
\address{$^\ddag$DAMTP, Centre for Mathematical Sciences, University
of Cambridge, Cambridge~CB3~0WA, UK}

\maketitle

\begin{abstract}

We investigate the effects on cosmological density perturbations
of dark radiation in a Randall-Sundrum~2 type brane-world. Dark
radiation in the background is limited by observational
constraints to be a small fraction of the radiation energy
density, but it has an interesting qualitative effect in the
radiation era. On large scales, it serves to slightly suppress the
radiation density perturbations at late times, while boosting the
perturbations in dark radiation. In a kinetic (stiff) era, the
suppression is much stronger, and drives the density perturbations
to zero.

\end{abstract}

% \vskip2pc]

\section{Introduction}
%%%%%%%%%%%%%%%%%%%%%%

In brane-world cosmology, which has emerged in the context of
recent developments in M~theory, the observable universe is a
1+3-dimensional ``brane" surface embedded in a 1+3+$d$-dimensional
``bulk" spacetime. Fields and particles in the non-gravitational
sector are confined to the brane, while gravity propagates in the
bulk. Simple phenomenological brane-world models are the
cosmological generalizations~\cite{bdel,unpert,sms,mwbh} of the
Randall-Sundrum~2 model~\cite{rs}, which has a self-gravitating
brane with Minkowski geometry embedded in an infinite part of
5-dimensional Anti de Sitter space (AdS$_5$). Gravity on the brane
is prevented from ``leaking" into the infinite extra dimension at
low energies via ``warping" of the metric by the negative bulk
cosmological constant,
% \be
$ \Lambda_5=-{6/ \ell^2},$
% \ee
where $\ell$ is the curvature radius of AdS$_5$. Modes of the 5D
graviton have an effective mass on the brane, since the projection
onto the brane of a null 5D graviton momentum vector is in general
a 4D timelike momentum vector. The massless mode arises when the
projection is null, and corresponds to the 4D graviton, which is
dominant at low energies. The massive Kaluza-Klein (KK) modes
produce corrections to the gravitational potential in the
weak-field static limit, which are $O(\ell^2/r^2)$~\cite{rs,gt}.
Table-top experiments currently impose an upper bound $\ell
\lesssim 0.1$~mm.

On the brane, the negative $\Lambda_5$ is offset by the positive
brane tension $\lambda$, and the effective cosmological constant
on the brane is
\begin{equation}\label{cc}
\Lambda=\frac{1}{2} (\Lambda_5+\kappa^2\lambda)\,,
\end{equation}
where $\kappa^2= {8\pi/ M_4^2}$ and $M_4\sim 10^{19}~$GeV is the
effective Planck scale on the brane. Fine-tuning can set
$\Lambda=0$. Because of the large extra dimension, the true
fundamental gravity scale can be as low as $\sim$~TeV in some
brane-world scenarios, but in generalized RS2 models it is higher:
\begin{equation}\label{scales}
M_5^3={M_4^2 \over \ell}\,,~ \ell<0.1~{\rm mm} ~\Rightarrow~
M_5>10^5~{\rm TeV}\,,~\lambda>(100~{\rm GeV})^4 \,.
\end{equation}
At high energies ($\rho\gg\lambda$) in the early universe, gravity
becomes 5-dimensional and there are significant corrections to
standard cosmological dynamics.

The unperturbed cosmological brane-world is a Friedmann brane in a
Schwarzschild-AdS$_5$ bulk~\cite{unpert} (where the expansion of
the universe can be interpreted as motion of the brane in the
bulk), satisfying the usual energy conservation equation but
modified Friedmann equations. For a spatially flat universe
without cosmological constant, these are
\begin{eqnarray}
H^2 &=& \frac{\kappa^2}{3} \rho\left(1+{\rho\over 2\lambda}\right)
 +{m\over a^4} \,, \label{e:friedmann1}\\
\dot H &=& - \frac{\kappa^2}{2} \rho(1+w)\left(1+{\rho\over
\lambda}\right)-2{m\over a^4}\,,
\end{eqnarray}
where $w=p/\rho$ and $m$ is a constant, proportional to the mass
of the bulk black hole. To avoid a naked singularity in the bulk,
we take $m\geq0$. The tidal, Coulomb effect of the 5-dimensional
black hole on the brane is an effective radiative ($\propto
a^{-4}$) term, the so-called ``dark radiation". Nucleosynthesis is
sensitive to the inclusion of additional relativistic energies not
thermally coupled to the radiation plasma, and so is the cosmic
microwave background. This places limits on the amount of dark
radiation compatible with observations~\cite{bm,lmsw,dr}
 \be\label{bbn}
\left({\cu\over \rho_{\rm rad}}\right)_{\rm nuc} \lesssim 0.03 \,,
 \ee
so that $\cu\ll \rho_{\rm rad}$ always, and the dark radiation is
negligible by the matter era. The quadratic correction terms
$\rho^2/\lambda$ are dominant at high energies, $\rho\gg\lambda$,
and negligible at low energies. Since $\lambda\gg 1~{\rm MeV}$,
the quadratic effects are negligible by nucleosynthesis. At late
times, the standard 4-dimensional evolution is therefore
recovered.

In the absence of dark radiation, the exact solution (with $\dot
w=0=m$) is~\cite{bdel}
 \be
a=\,\mbox{const}\,[t(t+t_\lambda)]^{1/3(w+1)}\,,~~t_\lambda
={M_4\over \sqrt{\pi\lambda}}<10^{-9}\,{\rm sec}\,.
 \ee
In the radiation era, $w={1\over3}$,  the exact solution with dark
radiation ($m\neq0$) is~\cite{bm}
 \be
a=\,\mbox{const}\,[t(t+t_\lambda)]^{1/4}\,,~~t_\lambda
={\sqrt{3}\,M_4\over 4\sqrt{\pi\lambda}(1+3m/\kappa^2\rho a^4)}\,.
 \ee
At late times, $t\gg t_\lambda$, we recover the standard solution
$a\propto t^{2/3(w+1)}$, whereas at early times (high energies),
$t\ll t_\lambda$, the evolution is very different: $a\propto
t^{1/3(w+1)}$.

The high-energy brane-world correction provides increased Hubble
damping in inflation; instead of $H\propto \sqrt{V}$ as in general
relativity, the modified Friedmann equation~(\ref{e:friedmann1})
shows that $H\propto V$ when $V\gg\lambda$. This means in
particular that slow-roll inflation is possible even for
potentials $V(\varphi)$ that would be too steep in standard
cosmology~\cite{mwbh,steep}. In addition, large-scale scalar
perturbations generated by high-energy slow-roll inflation have an
enhanced amplitude $A_{\rm s}$ compared with the standard general
relativity amplitude $(A_{\rm s})_{\rm gr}$~\cite{mwbh}:
 \be\label{inf}
A_{\rm s}^2 \approx
%\left[{64\pi\over 75 M_{4}^6}\,{V^3\over V'^2}\right]
\left(A_{\rm s}^2\right)_{\rm gr} \left( { {V\over\lambda
}}\right)^2.
 \ee

The background dynamics of brane-world cosmology are simple
because they are effectively 4-dimensional. Cosmological
perturbations introduce truly 5-dimensional degrees of freedom, so
that the 5-dimensional bulk perturbation equations must be solved
in order to solve for perturbations on the brane. These
5-dimensional equations are partial differential equations for the
3-dimensional Fourier modes, subject to complicated initial and
boundary conditions. However, on large scales, if one can neglect
gradient terms, the density or curvature perturbations can be
determined without knowing the 5-dimensional
solutions~\cite{mwbh,m1,lmsw}. The qualitative 5-dimensional
effects on large-scale density perturbations during inflation and
the transition to radiation-domination were investigated via a toy
model in~\cite{gm}. Here we generalise the results of~\cite{gm} to
include the dark radiation term $m/a^4$ in the background. The
presence of this term in the background can have a significant
effect on the perturbations.

\section{Field equations}

The effective field equations on the brane are derived from the 5D
field equations in the bulk,
$^{(5)}\!G_{AB}=-\Lambda_5\,^{(5)}\!g_{AB}$, by projecting the 5D
curvature (using the Gauss-Codazzi equations), and then imposing
the Darmois-Israel junction conditions at the brane (with
$Z_2$-symmetry)~\cite{sms}:
\begin{equation} \label{e:einstein1}
G_{ab} = - \Lambda g_{ab} + \kappa^2 T_{ab} +
6\frac{\kappa^2}{\lambda} {\cal S}_{ab} - {\cal E}_{ab}\;.
\end{equation}
Here ${\cal S}_{ab}\sim (T_{ab})^2$ is the high-energy correction
term, which is negligible for $\rho\ll\lambda$, while ${\cal
E}_{ab}$ is the projection of the bulk Weyl tensor, encoding
corrections from 5D graviton (KK) effects (and giving the dark
radiation in the background). From the brane-observer viewpoint,
the energy-momentum corrections in ${\cal S}_{ab}$ are local,
whereas the KK corrections in ${\cal E}_{ab}$ are
nonlocal~\cite{m1,mu,m2}, since they incorporate 5D gravity wave
modes. These nonlocal corrections cannot be determined purely from
data on the brane, and so the effective field equations are not a
closed system. One needs to supplement them by 5D equations
governing ${\cal E}_{ab}$, which are obtained from the 5D Einstein
and Bianchi equations~\cite{sms}.

The trace free ${\cal E}_{ab}$ contributes an effective KK energy
density $\rho^*$, pressure $\rho^*/3$, momentum density $q^*_a$
and anisotropic stress $\pi^*_{ab}$ on the brane,
 \be
-{1\over\kappa^2} {\cal E}_{ab} = {\cu}\left(u_a u_b+{ {1\over3}}
h_{ab}\right)+ {\cq_a} u_{b} + {\cq_b} u_{a}+\cp_{ab}\,,
 \ee
where $u^a$ is a physically determined 4-velocity on the brane and
$h_{ab}=g_{ab}+u_au_b$ projects into the comoving rest space at
each event. The brane ``feels" the bulk gravitational field
through these terms. In the background, $q^*_a=0=\pi^*_{ab}$,
since only the dark radiation term $\cu$ ($\propto ma^{-4}$) is
compatible with Friedmann symmetry. The KK momentum density
defines a velocity $v_a$ of the Weyl ``fluid" relative to $u^a$,
by $\cq_a=\cu v_a$.

For a perfect fluid or scalar field, we choose $u^a$ as the frame
in which there is no energy flux. Then the effective total energy
density and pressure are
\begin{eqnarray}
\rho^{\text{eff}} &=& \rho\left(1 +\frac{\rho}{2\lambda} +
\frac{\rho^*}{\rho} \right)\;, \\ \label{e:pressure1} p^{\text{eff
}} &=& \rho \left[w  + (1+2w)\frac{\rho}{2\lambda}
+\frac{\rho^*}{3\rho}\right]\;.
\end{eqnarray}
Energy-momentum conservation, $\nabla^b T_{ab}=0$, holds on the
brane. Together with the 4D Bianchi identity, this implies that
$\nabla^a{\cal E}_{ab}={6\kappa^2}\,\nabla^a{\cal
S}_{ab}\,/\lambda$, which shows qualitatively how 1+3 spacetime
variations in the matter-radiation on the brane can source KK
modes. These KK ``conservation equations" in linearized form are
\begin{eqnarray}
&& \dot{\rho}^*+{{4\over3}}\Theta{\cu}+\D^a{\cq_a}=0\,,
\label{nlc1}
\\&& \dot{q}^*_a+4H{\cq_a}
+{{1\over3}}\D_a{\cu}+{{4\over3}}{\cu}A_a +\D^b{\cp_{ab}} =-{
(\rho+p)\over\lambda}\D_a \rho\,,\label{nlc2}
\end{eqnarray}
where $\Theta$ is the volume expansion rate ($=3H$ in the
background), $A_a$ is the 4-acceleration, and $\D_a$ is the
covariant spatial derivative in the comoving rest space. Spatial
inhomogeneity ($\D_a \rho\neq0$) is a source for KK modes.
Qualitatively and geometrically this can be understood as
follows~\cite{m1,mu}: the non-uniform 5D gravitational field
generated by inhomogeneous 4D matter-radiation contributes to the
5D Weyl tensor, which nonlocally ``backreacts" on the brane via
its projection ${\cal E}_{ab}$. Equation~(\ref{nlc2}) shows that
the source term is suppressed at low energies, and during quasi-de
Sitter inflation on the brane.

Equations~(\ref{nlc1}) and (\ref{nlc2}) are propagation equations
for $\cu$ and $\cq_a$. There is no propagation equation on the
brane for $\cp_{ab}$, so that one cannot determine the KK modes
purely from data on the brane. In the 1+3-covariant description of
braneworld perturbations~\cite{m1,l1,bd,l2}, the KK anisotropic
stress $\cp_{ab}$ is isolated as the term that must be determined
from 5D equations. Once $\cp_{ab}$ is determined in this way, the
1+3 perturbation equations on the brane form a closed system. The
KK terms act as source terms modifying the standard general
relativity perturbation equations, together with the local
high-energy corrections.

\section{ Density perturbation equations}
%%%%%%%%%%%%%%%%%%%%%%%%%%%%%%%%%%%%

We define density and expansion (velocity) perturbation scalars,
as in general relativity~\cite{bde},
 \be
\Delta={a^2\over\rho}\D^2\rho\,,~Z=a^2\D^2\Theta\,,
 \ee
and then define dimensionless KK perturbation scalars~\cite{m1},
 \be
{   U}={a^2\over\rho}\D^2{\cu}\,,~{   Q}={a\over\rho} \D^2
{\cq}\,,~{  \Pi }={1\over \rho}\D^2{\cp}\,,
 \ee
where the scalar potentials $\cq$ and $\cp$ are defined by
$\cq_a=\D_a\cq$, $\cp_{ab}=(\D_a\D_b-{1\over3}h_{ab}\D^2)\cp$. The
dark radiation fluctuation $U$ is present even if there is no dark
radiation in the background ($\cu=0$). It leads to an isocurvature
(non-adiabatic) mode, even when the matter perturbations are
assumed adiabatic~\cite{gm}. We define the total effective
dimensionless entropy $S^{\rm eff}$ via the non-adiabatic part of
the effective pressure as~\cite{bde}
 \be
p^{\rm eff}\,S^{\rm eff}=a^2\D^2 p^{\rm eff}-c_{\rm
eff}^2a^2\,\D^2\rho^{\rm \,eff}\,,
 \ee
where $c_{\rm eff}^2=\dot{p}^{\rm eff}/\dot{\rho}^{\rm eff}$. Then
\begin{eqnarray}
S^{\rm eff}={\left[3c_{\rm s}^2 - {1} +\left({2}+3w+3c_{\rm
s}^2\right){ {\rho/\lambda}}\right]\over
[(1+w)(1+\rho/\lambda)+4\cu/3\rho] [3w+3(1+2w)\rho/
2\lambda+\cu/\rho]}\, \left[ {4\over 3}{\cu\over \rho}\,\Delta
-(1+w) U \right]\,,\label{ent}
\end{eqnarray}
where $c_{\rm s}^2=\dot{p}/\dot{\rho}$. If $\cu=0$ in the
background, then $U$ is an isocurvature mode: $S^{\rm eff}\propto
U$. If $\cu\neq0$ in the background, then the weighted difference
between $U$ and $\Delta$ determines the isocurvature mode: $S^{\rm
eff}\propto (4\cu/ 3\rho)\Delta -(1+w)U$. At very high energies,
$\rho\gg\lambda$, the entropy is suppressed by the factor
$\lambda/\rho$. At low energies in the radiation era ($c_{\rm
s}^2={1\over3}$) it is suppressed by the factor $\rho/\lambda$.

The covariant density perturbation equations on the
brane~\cite{m1} reduce to
\begin{eqnarray}
\dot{\Delta} &=&3wH\Delta-(1+w)Z\,, \label{p1}\\  \dot{Z}
&=&-2HZ-\left({c_{\rm s}^2\over 1+w}\right)
\D^2\Delta-\kappa^2\rho {   U}-{{1\over2}}\kappa^2 \rho\left[1+
(4+3w){ {\rho\over\lambda}}- \left({4c_{\rm s}^2\over
1+w}\right){\cu\over\rho}\right] \Delta \label{p5}\,,\\  {
\dot{U}} &=& (3w-1)H{   U} + \left({4c_{\rm s}^2\over
1+w}\right)\left({{\cu }\over\rho}\right) H\Delta -\left({4{\cu
}\over3\rho}\right) Z-a\D^2{ Q}\,, \label{p6}\\  {   \dot{Q}}
&=&(3w-1)H{ Q}-{1\over3a}{ U}-{{2\over3}} a{ \D^2\Pi}+{1\over3
a}\left[ \left({4c_{\rm s}^2\over
1+w}\right){{\cu}\over\rho}-3(1+w) {
{\rho\over\lambda}}\right]\Delta\,,\label{p4}
\end{eqnarray}
where $\dot f\equiv u^a\partial_a f$, and we have corrected the
minor errors in the equations given in~\cite{gm}. The KK
anisotropic stress term $\Pi$ occurs only via its Laplacian, ${
\D^2\Pi}$. If we can neglect this term on large scales, then the
system of density perturbation equations closes on super-Hubble
scales~\cite{m1}. An equivalent statement applies to the
large-scale curvature perturbations~\cite{lmsw}. We note that in
the low-energy approximation scheme used by Koyama~\cite{koy},
 \be
\Pi \propto \D^2U\,,
 \ee
so that in particular $\Pi\to0$ on large scales.

On large scales in the braneworld, neglecting gradient terms,
density perturbations are then described by the set
$\{\Delta,Z,U,Q\}$, where $Q$ decouples from the other variables.
KK effects introduce two new isocurvature modes on large scales,
associated with $U$ and $Q$~\cite{gm,l1}.

We can find a first integral of the large-scale system by
following the same approach used in general relativity~\cite{bde}.
A covariant local curvature perturbation is defined by the
gradient of $R^\perp$, the 3-Ricci scalar for $u^a$-observers.
This scalar is given by the Codazzi equation on the
brane~\cite{m2},
 \be
R^\perp=-{2\over 3}\Theta^2+2\kappa^2\rho^{\rm eff}\,.
 \ee
Then the dimensionless curvature perturbation scalar is defined by
$C=a^4\D^2R^\perp$. This gives
 \be
C=-4a^2HZ+2\kappa^2a^2\rho\left(1+{\rho\over \lambda}
\right)\Delta + 2\kappa^2a^2\rho U\,.
 \ee
Using the large-scale equations~(\ref{p1})--(\ref{p6}), we find
that $C$ is locally conserved (i.e., constant along fundamental
worldlines) on large scales,
 \be \label{c}
C=C_0\,,~~\dot C_0=0\,.
 \ee
(The minor errors in~\cite{gm} led to the erroneous conclusion
that $C$ was not in general locally conserved on the brane.)

Thus $C$ leads to a first integral for the large-scale system. We
also replace $\Delta$ by the dimensionless perturbation variable
 \be\label{phi3}
\Phi=\kappa^2a^2\rho \Delta\,.
 \ee
In general relativity, $\Phi$ is the covariant analogue of the
Bardeen metric perturbation variable $\Phi_H$~\cite{bde}. In the
brane-world, high-energy and KK effects mean that the analogue of
$\Phi_H$ is a complicated generalisation of Eq.~(\ref{phi3})
(see~\cite{l1}). The key problem in trying to use the generalised
variable is that it contains $\Pi$, which remains undetermined.
(This is the same reason that the large-scale Sachs-Wolfe effect
on the brane cannot be determined from the density or curvature
perturbations~\cite{lmsw}.) In any case, $\Phi$ as defined above
leads to useful simplification of the perturbation equations.

The coupled system of Eqs.~(\ref{p1})--(\ref{p6}), which is closed
on large scales (neglecting gradient terms), reduces via
Eqs.~(\ref{c}) and (\ref{phi3}) to the system
\begin{eqnarray}
\dot{\Phi}&=& -H\left[1+(1+w){\kappa^2\rho\over 2H^2}\left(1+
{\rho\over \lambda}\right)\right]\Phi  -
\left[(1+w){a^2\kappa^4\rho^2\over 2 H}\right]U +\left[(1+w)
{\kappa^2 \rho\over 4H}\right]C_0\,, \label{p1'}\\ \dot{U} &=&
-H\left[1-3w+{2\kappa^2{\cu}\over 3H^2}\right]U -{2 {\cu}\over 3
a^2 H\rho}\left[1+{\rho\over\lambda} - {6 c_{\rm s}^2H^2\over
(1+w)\kappa^2\rho}\right]\Phi+ \left[{{\cu}\over 3a^2H\rho}\right]
C_0\,. \label{p3}
\end{eqnarray}
This closed system for $\Phi, U$ generalises the equations given
in~\cite{gm} to the case $\cu\neq0$. Once $\Phi$ and $U$ are
solved for, $Q$ is directly determined by Eq.~(\ref{p4}) with
$\D^2\Pi$ set to zero.

\section{Solutions}

The closed system for $\Phi$ and $U$ may be rewritten on using the
Friedmann equation~(\ref{e:friedmann1}) and changing the time
variable to the number of e-folds, $t\to N=\ln(a/a_0)$. This
yields
\begin{eqnarray}
{\Phi}'&=& -\left[1+{3(1+w)(1+\rho/\lambda) \over
2(1+\rho/2\lambda +\cu/\rho)}\right]\Phi - \left[ {3(1+w)\over
2(1+\rho/2\lambda +\cu/\rho)}\, {a^2\kappa^2\rho}\right]U +\left[
{3(1+w) \over 4(1+\rho/2\lambda +\cu/\rho)}\right]C_0\,,
\label{p1''}\\ {U}' &=& -\left[1-3w+{2{\cu}/\rho\over
(1+\rho/2\lambda +\cu/\rho)}\right]U -\left[{\cu\over\rho}\left\{
{(1+\rho/\lambda) \over (1+\rho/2\lambda +\cu/\rho)}  - {2 c_{\rm
s}^2\over (1+w)}\right\} {2\over a^2\kappa^2\rho}\right]\Phi
\nonumber\\ &&~{}+ \left[{\cu\over\rho}\,{1 \over (1+\rho/2\lambda
+\cu/\rho)}\,{1\over a^2\kappa^2 \rho}\right] C_0\,, \label{p3'}
\end{eqnarray}
where $f'=df/dN$ and
 \be\label{beta}
a^2\kappa^2\rho=\beta \exp-\int (1+3w)dN\,,~\beta= a_0^2\kappa^2
\rho_0\,.
 \ee
Here $\beta$ is dimensionless and can be given an arbitrary
positive value by suitable choice of $a_0$.

For comparison, in the general relativity case
($\rho/\lambda=0=\cu=U$) these two equations reduce to the single
equation
 \be
\Phi_{\rm gr}'=-{1\over2}(5+3w)\Phi_{\rm gr}+{3\over 4}(1+w)C_0\,.
 \ee
If $w$ is constant, then the non-decaying attractor solution is
 \be\label{gr}
\Phi_{\rm gr}={3(1+w) \over 2(5+3w)}\, C_0\,.
 \ee
Deviations from general relativity arise from early-universe
high-energy effects (when $\rho\gg\lambda$), and from the KK
effects encoded in $\cu$ and $U$. At late times, the high-energy
corrections become negligible. The background dark radiation
decays like ordinary radiation ($\cu\propto a^{-4}\propto\rho_{\rm
rad} $), and is negligible at late times by Eq.~(\ref{bbn}).
However, it can leave a signature that survives at late times, as
we show below.

\subsection{No dark radiation in the background}

Equation~(\ref{p3'}) is much simpler in the case when there is no
dark radiation in the background ($\cu=0$), i.e. when there is no
bulk black hole. (Note also that the errors in~\cite{gm} do not
affect this case, so that the solutions given there are correct.)
When $\cu=0$ we get
 \be
U=U_0\exp-\int(1-3w)dN\,.
 \ee
Since $\cu=0$, it follows from Eq.~(\ref{ent}) that $U$ is an
isocurvature mode. This isocurvature mode associated with dark
radiation perturbations therefore decays for all equations of
state less stiff than radiation, i.e. for $w<{1\over3}$.

If $w$ is constant, then Eq.~(\ref{p1''}) becomes
 \be \label{phi}
{\Phi}'= -\left[{(5+3w)+(4+3w)\rho/\lambda \over
2+\rho/\lambda}\right]\Phi - \left[ {3(1+w)\over
2(1+\rho/2\lambda)}\right]e^{-2N}\,\tilde{U}_0 +\left[ {3(1+w)
\over 4(1+\rho/2\lambda)}\right]C_0\,,
 \ee
where $\tilde{U}_0=\beta U_0$. The $U$ contribution to $\Phi$ is
suppressed at low energies, even if $w>{1\over 3}$ (when $U$
grows). Thus the general relativity solution Eq.~(\ref{gr}) for
constant $w$ is a low-energy attractor for $\cu=0$:
 \be\label{low}
\Phi_{\rm low} \approx {3(1+w) \over 2(5+3w)}\, C_0\,.
 \ee
Thus there is no surviving imprint in $\Phi$ of the dark radiation
perturbations (the isocurvature mode) when $\cu=0$ in the
background.

At high energies/ early times, Eq.~(\ref{phi}) becomes
 \be \label{phi2}
{\Phi}'\approx -(4+3w)\Phi - {\lambda\over\rho_0}\left[ 3
(1+w)\right]e^{(1+3w)N}\,\tilde{U}_0 +{\lambda \over \rho_0 }
\left[ {3\over2}(1+w) \right]e^{3(1+w)N}C_0\,,
 \ee
where $\lambda\ll\rho_0$. Thus in the early universe, the
contribution of the entropy mode $U$ is sensitive to the equation
of state. This contribution decays for $w<-{1\over3}$, including
the case of inflation. For $w\geq -{1\over3}$, Eq.~(\ref{phi2})
shows that the non-decaying solution is
 \bea
\Phi_{\rm high} & \approx & {\lambda \over \rho_0}\left[{3(1+w)
\over 2(7+6w)}\, C_0\right]e^{3(1+w)N}-{\lambda \over \rho_0}
\left[{3\beta (1+w) \over (5+6w)}\, U_0\right]e^{(1+3w)N}\,,
 \eea
so that in the very high energy limit,
 \be\label{vhe}
\Phi_{\rm high} \to  {3\over 2}{\lambda \over \rho_0}(1+w) \left[
{C_0 \over 7+6w}-{2\tilde{U}_0 \over 5+6w} \right]\,.
 \ee
Hence, for constant $w$ ($\geq -{1\over 3}$), high-energy effects
initially suppress $\Phi$ (by the factor $\lambda/\rho_0$), before
it starts to grow. This growth leads to the late-time general
relativity value, Eq.~(\ref{low}). By contrast, in general
relativity, $\Phi$ remains constant when $w$ is constant.

The most relevant case is $w={1\over3}$. If inflation takes place
on the brane, reheating will initiate a radiation era. Provided
the reheat temperature is high enough, the radiation era will
start at high energies. Large-scale perturbations $\Phi$ will be
amplified according to Eqs.~(\ref{vhe}) and (\ref{low}) with
$w={1\over3}$.

\begin{figure}[!bth]\label{inf2}
\begin{center}
\includegraphics[scale=1]{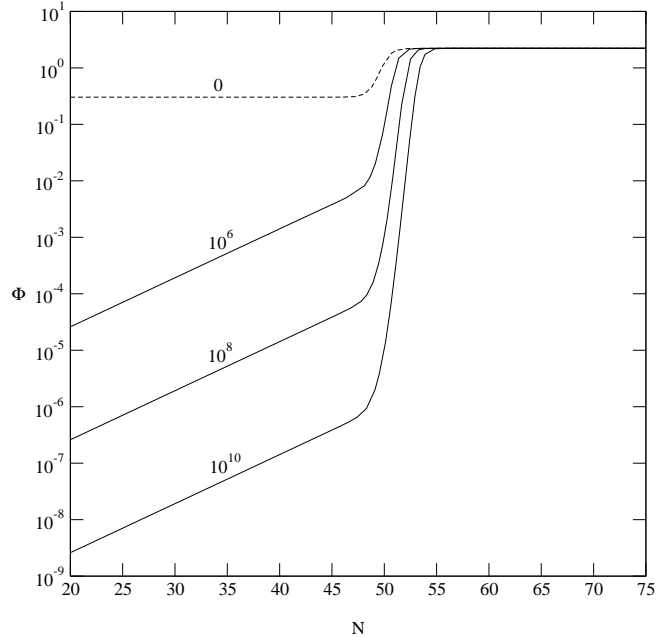}
\caption{The evolution of $\Phi$ for a mode that is well beyond
the Hubble radius at $N =0$, about 50 e-folds before inflation
ends, and remains super-Hubble through the radiation era. A smooth
transition from inflation to radiation is modelled by
$w={1\over2}[({4\over3}-\epsilon)\tanh(N-50)-({2\over3}
-\epsilon)]$, where $\epsilon$ is a small positive parameter,
chosen as $\epsilon=1/15$ in the plot. (The parameter $\beta$ in
Eq.~(\ref{beta}) is set to 1). Labels on the curves give the value
of $\rho_0/\lambda$; the general relativity solution is the dashed
curve. }
\end{center}
\end{figure}

For slow-roll inflation, $w$ is not constant, but close to -1 and
slowly increasing, i.e., $1+w\sim\epsilon$, with $0<\epsilon \ll
1$ and $H^{-1}|\dot\epsilon|=|\epsilon'|\ll 1$. Then
Eq.~(\ref{phi2}) leads to
 \be
\Phi_{\rm high} \sim {3\over 2}\epsilon {\lambda \over \rho_0}C_0
e^{3\epsilon N}\,,
 \ee
so that $\Phi$ has a slowly growing mode during high-energy
slow-roll inflation. This is different from general relativity,
where $\Phi$ is constant during slow-roll inflation. Thus more
amplification of $\Phi$ can be achieved than in general
relativity, consistent with Eq.~(\ref{inf}). The evolution of
$\Phi$ is illustrated for a toy model of inflation-to-radiation in
Fig.~1 (from~\cite{gm}). The early (growing) and late time
(constant) attractor solutions are seen explicitly in the plots.

\subsection{Nonzero dark radiation in the background}

The presence of dark radiation in the background has interesting
qualitative implications, even though the quantitative constraints
from observations on the amount of dark radiation are quite
severe. The general relativity solution, Eq.~(\ref{gr}) is no
longer the late-time attractor when $\cu\neq0$.

The dark radiation factor, for constant $w$, behaves as
 \be\label{cur}
{\cu\over\rho} =\alpha e^{(3w-1)N}\,,
~~\alpha={\cu_0\over\rho_0}\,,
 \ee
where $\alpha$ is a dimensionless constant. For equations of state
less stiff than radiation ($w<{1\over3}$), the dark radiation
factor redshifts away. If $w>{1\over3}$, then this factor grows
with expansion. In the radiation era, $w={1\over3}$, the dark
radiation factor is constant.

If there was a period of inflation on the brane before the
radiation era, then dark radiation would be redshifted to
negligible levels. However, the reheating era that creates the
radiation, could at the same time create dark radiation via 5D
graviton emission in high-energy interactions~\cite{lsr}. The
fraction of dark radiation is limited by observational
constraints, Eq.~(\ref{bbn}), so that
 \be
\alpha \lesssim 0.03\,.
 \ee
We can therefore perform a perturbative solution of
Eqs.~(\ref{p1''}) and (\ref{p3'}), up to first order in $\alpha$.
The zero-order solutions are given in the previous subsection.

We find that the non-decaying solutions for radiation in the
low-energy regime are
 \bea
\Phi_{\rm low} & \approx & {C_0\over 3}(1-\alpha)\,,\label{pdr}\\
\tilde{U}_{\rm low} & \approx & \tilde{U}_0+ \alpha
{C_0\over3}\,e^{2N}\,.\label{udr}
 \eea
It follows that the first-order correction to $\tilde{U}$ (where
$\tilde{U}=\tilde{U}^{(0)}+\tilde{U}^{(1)}$) behaves like the
large-scale density perturbations for radiation in general
relativity:
 \be
\tilde{U}_{\rm low}^{(1)} \propto \Delta \propto a^2\,.
 \ee
In the high-energy regime, the non-decaying solutions  are
 \bea
\Phi_{\rm high} & \approx & {\lambda\over \rho_0}\left[{2\over9}
C_0e^{4N} -{4\over7}\tilde{U}_0e^{2N}\right]+4\alpha
\left({\lambda\over
\rho_0}\right)^2\left[\left({2\over63}C_0-{4\over49}
\tilde{U}_0\right)e^{2N} + {4\over77}\tilde{U}_0e^{6N}-{2\over
117} C_0 e^{8N}\right]\,,\label{hdrp} \\ \tilde{U}_{\rm high} &
\approx & \tilde{U}_0+ \alpha {\lambda\over \rho_0}\left[{2\over9}
C_0 \left(e^{6N}-1\right)-{4\over7}\tilde{U}_0
\left(e^{4N}-1\right) \right] \,.\label{hdru}
 \eea
In the very high energy limit,
 \bea
\Phi_{\rm high} & \to & {\lambda\over \rho_0}\left[{2\over9} C_0
-{4\over7}\tilde{U}_0\right]+16\alpha \left({\lambda\over
\rho_0}\right)^2\left[{C_0\over273}-{4\tilde{U}_0 \over
539}\right] \,,\label{vhe2}\\ \tilde{U}_{\rm high} & \to &
\tilde{U}_0\,.
 \eea
Comparing Eq.~(\ref{vhe2}) with Eq.~(\ref{vhe}) shows that the
high energy limit with dark radiation differs only negligibly from
the no-dark-radiation limit.

\begin{figure}[!bth]\label{phirad}
\begin{center}
\includegraphics[scale=0.5]{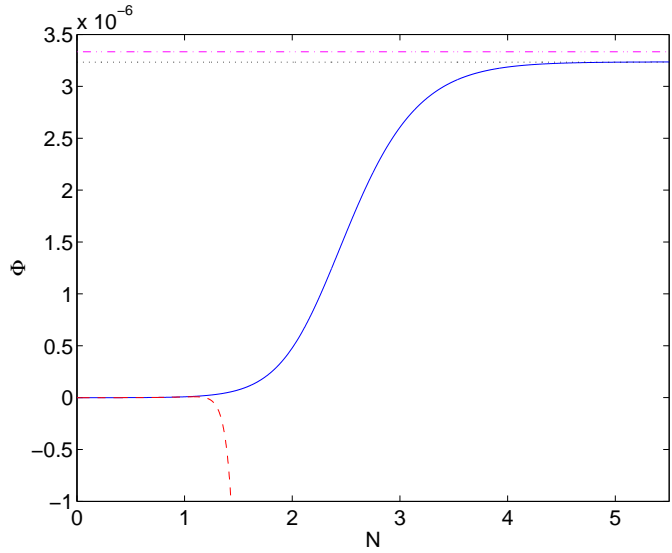}
\caption{The evolution of $\Phi$ on very large scales in the
radiation era, with $\rho_0/\lambda= 10^4$ and dark radiation
factor $\alpha=0.03$. Initial conditions are given by
$C_0=10^{-5}=\tilde{U}_0$, and $\Phi$ is initially set to its
high-energy attractor, Eq.~(\ref{vhe2}). The solid curve is the
numerical integration. The dashed curve is the high-energy
approximation, Eq.~(\ref{hdrp}). The dotted curve is the
low-energy approximation, Eq.~(\ref{pdr}), and the dot-dashed
curve is the low-energy approximation without dark radiation
($\alpha=0$).
 }
\end{center}
\end{figure}

This analysis shows that $\Phi$ is initially suppressed, then
begins to grow, as in the no-dark-radiation case, eventually
reaching an attractor which is less than the no-dark-radiation
attractor. The role of dark radiation is to suppress, by the small
fractional amount $\alpha$, the final value of $\Phi$, given in
Eq.~(\ref{pdr}). The dark radiation perturbation $U$ starts from a
constant value at very high energies, and then grows at late times
like the density perturbations in radiation. Figures~2 and 3
confirm the qualitative analysis.

\begin{figure}[!bth]\label{urad}
\begin{center}
\includegraphics[scale=0.5]{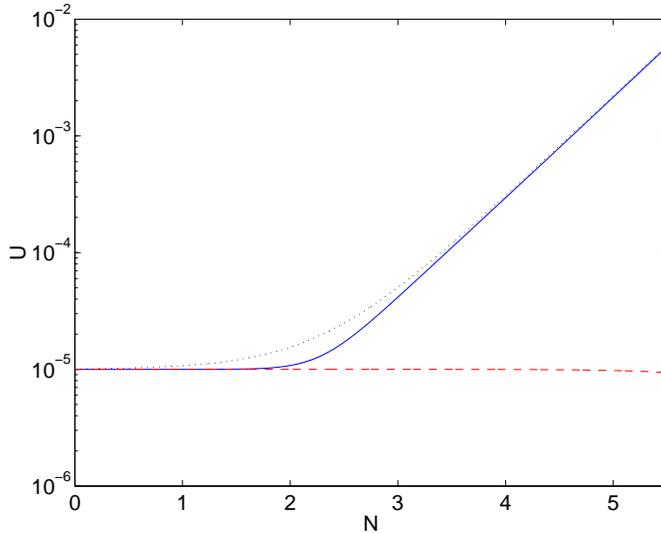}
\caption{The evolution of $U$, with the same conditions as in
Fig.~2. The dashed curve is the high-energy approximation,
Eq.~(\ref{hdru}), and the dotted curve is the low-energy
approximation, Eq.~(\ref{udr}).
 }
\end{center}
\end{figure}

The effective total entropy, Eq.~(\ref{ent}), has the following
behaviour, to $O(\alpha)$, in the radiation era:
 \bea
&& S^{\rm eff}_{\rm high} \to -{8\over5}{\lambda\over
\rho_0}\left[ U_0 -\alpha{\lambda\over \rho_0}\left( {2C_0\over
9\beta}-{4\over7}U_0\right)\right]\,,\\ && S^{\rm eff}_{\rm low}
\approx -4{\rho_0\over\lambda}e^{-4N}(1-2\alpha)U_0\,.
 \eea
At high energies $|S^{\rm eff}|$ is suppressed by the factor
$\lambda/\rho_0$, and the $\alpha$-correction is negligible. At
low energies, the large factor $\rho_0/\lambda$ is overwhelmed by
the redshift factor $e^{-4N}$.

Without dark radiation in the background, $U$ in the radiation era
remains constant. When there is dark radiation in the background,
$U$ grows. The growth of $U$ leads to a decrease in the final
value of $\Phi$.

An extreme case of loss of power in $\Phi$ occurs when the dark
radiation redshifts more slowly than the matter, so that it
eventually dominates over the matter. This happens when the matter
equation of state is stiffer than radiation, $w> {1\over3}$. For
example, for stiff matter, $w=1$, we have
 \be
{\cu\over\rho} \propto a^2\,,
 \ee
by Eq.~(\ref{cur}). As a result, the decrease in $\Phi$ becomes so
pronounced that $\Phi$ is driven to 0 at late times. By
Eqs.~(\ref{p1''}) and (\ref{p3'}), we find that
 \be
U_{\rm low}\propto e^{4N}\,,~~ \Phi_{\rm low}\propto e^{-2N}\,,
 \ee
at late times. The suppression of $\Phi$ is confirmed by numerical
integration, as shown in Fig.~4.

\begin{figure}[!bth]\label{phistiff}
\begin{center}
\includegraphics[scale=0.5]{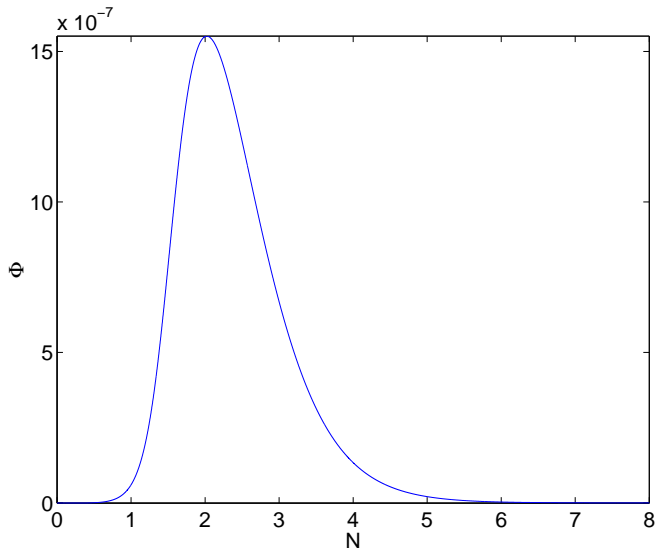}
\caption{The evolution of $\Phi$ on large scales in a stiff
(kinetic) era (with parameters and initial conditions as in
Figs.~2 and 3). }
\end{center}
\end{figure}

This example could be relevant to the steep inflation scenario on
the braneworld. High-energy effects mean that steep potentials can
drive inflation on the brane when they would not give inflation in
general relativity. Typically, this inflation ends in a kinetic
regime with $w\approx1$, and with particle production being
gravitational~\cite{steep}. If dark radiation can be produced by
high-energy interactions, then its presence serves to suppress
$\Phi$ on large-scales, as in Fig.~4.

\section{Conclusion}
%%%%%%%%%%%%%%%%%%%%

We have investigated the effects of dark radiation in the
background on large-scale density perturbations (neglecting
gradient terms) in a Randall-Sundrum~2 type brane-world. Dark
radiation significantly complicates the perturbation equations.
Since it is limited to be a small fraction of the radiation energy
density, its effects can be qualitatively analysed via a
perturbative approach. This approach, confirmed by our numerical
simulations, shows that in the radiation era, the large-scale
density perturbations are suppressed at late times by a small
amount. At the same time, the large-scale perturbations in the
dark radiation itself grow at late times. By contrast these
perturbations have constant amplitude in a universe with no dark
radiation in the background. The suppression of density
perturbations becomes strong when the matter has a stiff
(kinetic-dominated) equation of state.

\[ \]
{\bf Acknowledgments}

We thank Peter Dunsby for pointing out the minor errors
in~\cite{gm}, which are corrected in Eqs.~(\ref{p5})--(\ref{p4}).
BG is supported by the British-Thai Scholarship Scheme and a Royal
Thai Government Scholarship. CG and RM are supported by PPARC.


\begin{references}

\bibitem{bdel}
P. Binetruy, C. Deffayet, U. Ellwanger and D. Langlois, Phys.
Lett. {\bf B477}, 285 (2000).

\bibitem{unpert}
P. Kraus, JHEP {\bf 12}, 011 (1999);\\ S. Mukohyama, Phys. Lett.
{\bf B473}, 241 (2000);\\ D. Ida, JHEP {\bf 09}, 014 (2000);\\ S.
Mukohyama, T. Shiromizu and K. Maeda, Phys. Rev. D {\bf 61},
024028 (2000);\\  E.E. Flanagan, S.-H. Henry Tye and I. Wasserman,
Phys. Rev. D {\bf 62}, 044039 (2000);\\ P. Bowcock, C. Charmousis
and R. Gregory, Class. Quantum Grav. {\bf 17}, 4745 (2000).

\bibitem{sms} T. Shiromizu, K. Maeda and M. Sasaki,
Phys. Rev. D {\bf 62}, 024012 (2000).

\bibitem{mwbh}
R. Maartens, D. Wands, B.A. Bassett and I.P.C. Heard, Phys. Rev. D
{\bf 62}, 041301 (2000).

\bibitem{rs} L. Randall and R. Sundrum, Phys. Rev. Lett. {\bf 83},
4690 (1999).

\bibitem{gt}
J. Garriga and T. Tanaka, Phys. Rev. Lett. {\bf 84}, 2778 (2000).

\bibitem{bm}
J.D. Barrow and R. Maartens, Phys. Lett. {\bf B532}, 153 (2002).

\bibitem{lmsw}
D. Langlois, R. Maartens, M. Sasaki and D. Wands, Phys. Rev. D
{\bf 63}, 084009 (2001).

\bibitem{dr}
K. Ichiki, M. Yahiro, T. Kajino, M. Orito and G.J. Mathews, Phys.
Rev. D {\bf 66}, 043521 (2002);\\ J.D. Bratt, A.C. Gault, R.J.
Scherrer and T.P. Walker, Phys. Lett. {\bf B546}, 19 (2002).

\bibitem{steep}
E.J. Copeland, A.R. Liddle and J.E. Lidsey, Phys. Rev. D {\bf 64},
023509 (2001);\\  A. S. Majumdar, Phys. Rev. D {\bf 64}, 083503
(2001);\\ G. Huey and J.E. Lidsey, Phys. Lett. {\bf B514}, 217
(2001);\\ V. Sahni, M. Sami and T. Souradeep, Phys. Rev. D {\bf
65}, 023518 (2002);\\ N.J. Nunes and E.J. Copeland, Phys. Rev. D
{\bf 66}, 043524 (2002);\\ A.R. Liddle, L.A. Urena-Lopez,
astro-ph/0302054.

\bibitem{m1}
R. Maartens, Phys. Rev. D {\bf 62}, 084023 (2000).

\bibitem{gm}
C. Gordon and R. Maartens, Phys. Rev. D {\bf 63}, 044022 (2001).

\bibitem{mu}
S. Mukohyama, Phys. Rev. D {\bf 62}, 084015 (2000); \\ S.
Mukohyama, Phys. Rev. D {\bf 64}, 064006 (2001).

\bibitem{m2}
R. Maartens, gr-qc/0101059.

\bibitem{l1}
B. Leong, P.K.S. Dunsby, A.D. Challinor and A.N. Lasenby, Phys.
Rev. D {\bf 65}, 104012 (2002).

\bibitem{bd}
M. Bruni and P.K.S. Dunsby, Phys. Rev. D {\bf 66}, 101301 (2002).

\bibitem{l2}
B. Leong, A.D. Challinor, R. Maartens and A.N. Lasenby, Phys. Rev.
D {\bf 66}, 104010 (2002).

\bibitem{bde}
M. Bruni, P.K.S. Dunsby and G.F.R. Ellis, Astrophys. J. {\bf 395},
34 (1992).

\bibitem{koy}
K. Koyama, astro-ph/0303108.

\bibitem{lsr}
D. Langlois, L. Sorbo and M. Rodriguez-Martinez, Phys. Rev. Lett.
{\bf 89}, 171301 (2002).


\end{references}
\end{document}